\documentstyle[11pt,newpasp,twoside,epsf,psfig]{article}
\def\edcomment#1{\iffalse\marginpar{\raggedright\sl#1\/}\else\relax\fi}
\reversemarginpar

\begin{document}
\title{A chemical evolution model for galaxy clusters}
 \author{L.~Portinari}
\affil{Theoretical Astrophysics Center, Juliane Maries Vej 30,\\
DK-2100 Copenhagen \O, Denmark}
\author{A.~Moretti and C.~Chiosi}
\affil{Dipartimento di Astronomia, Vicolo dell'Osservatorio 2, \\
I-35122 Padova, Italy}

\begin{abstract}
We develop a toy--model for the chemical evolution of the intra--cluster 
medium, polluted by the galactic winds from elliptical galaxies.
The model follows the ``galaxy formation history'' of cluster 
galaxies, constrained by the observed luminosity function.
\end{abstract}
\vspace{-5mm}
\section{Introduction}
To account for the large amount of metals observed in the intra--cluster
medium (ICM), some non-standard stellar Initial Mass Function (IMF) 
has been often invoked for cluster ellipticals, 
such as a more top--heavy IMF than the Salpeter one (Matteucci \& Gibson 1995;
Gibson \& Matteucci 1997ab; Loewenstein \& Mushotzky 1996), or 
a bimodal IMF with an early generation of massive stars 
(Arnaud et~al.\ 1992; Elbaz, Arnaud \& Vangioni-Flam 1995);
see also the review by Matteucci (this conference).
A non--standard IMF in ellipticals has been suggested 
also on the base of other arguments:
a top--heavy IMF best reproduces their photometric properties
(Arimoto \& Yoshii 1987),
and sistematic variations of the IMF in ellipticals of increasing mass might 
explain the observed trend $M/L \propto L$
(Larson 1986, Renzini \& Ciotti 1993, Zepf \& Silk 1996).

Chiosi et~al.\ (1998) developed chemo-spectrophotometric models for
elliptical galaxies with galactic winds, adopting the variable IMF 
by Padoan, Nordlund \& Jones (1997, hereinafter PNJ) which is naturally skewed
toward more massive stars in the early galactic phases, in more massive 
galaxies and for higher redshifts of formation (see also Chiosi 2000).
These galactic models were successful at reproducing a number of 
spectro--photometric features of observed ellipticals;
now, an immediate question is: what do these galactic models predict for
the pollution of the ICM through galactic winds (GWs)?

\section{Galactic wind ejecta: PNJ vs.\ Salpeter IMF}
To address this issue, Chiosi (2000) calculated multi--zone chemical models
of elliptical galaxies with the PNJ IMF, together with models with 
the standard Salpeter IMF for the sake of comparison.
Before discussing the resulting global chemical evolution of the ICM,
let's inspect the different GW ejecta of the model ellipticals
when the two IMFs are adopted in turn.

\newpage
\begin{figure}[t]
{\centering 
\leavevmode
\psfig{file=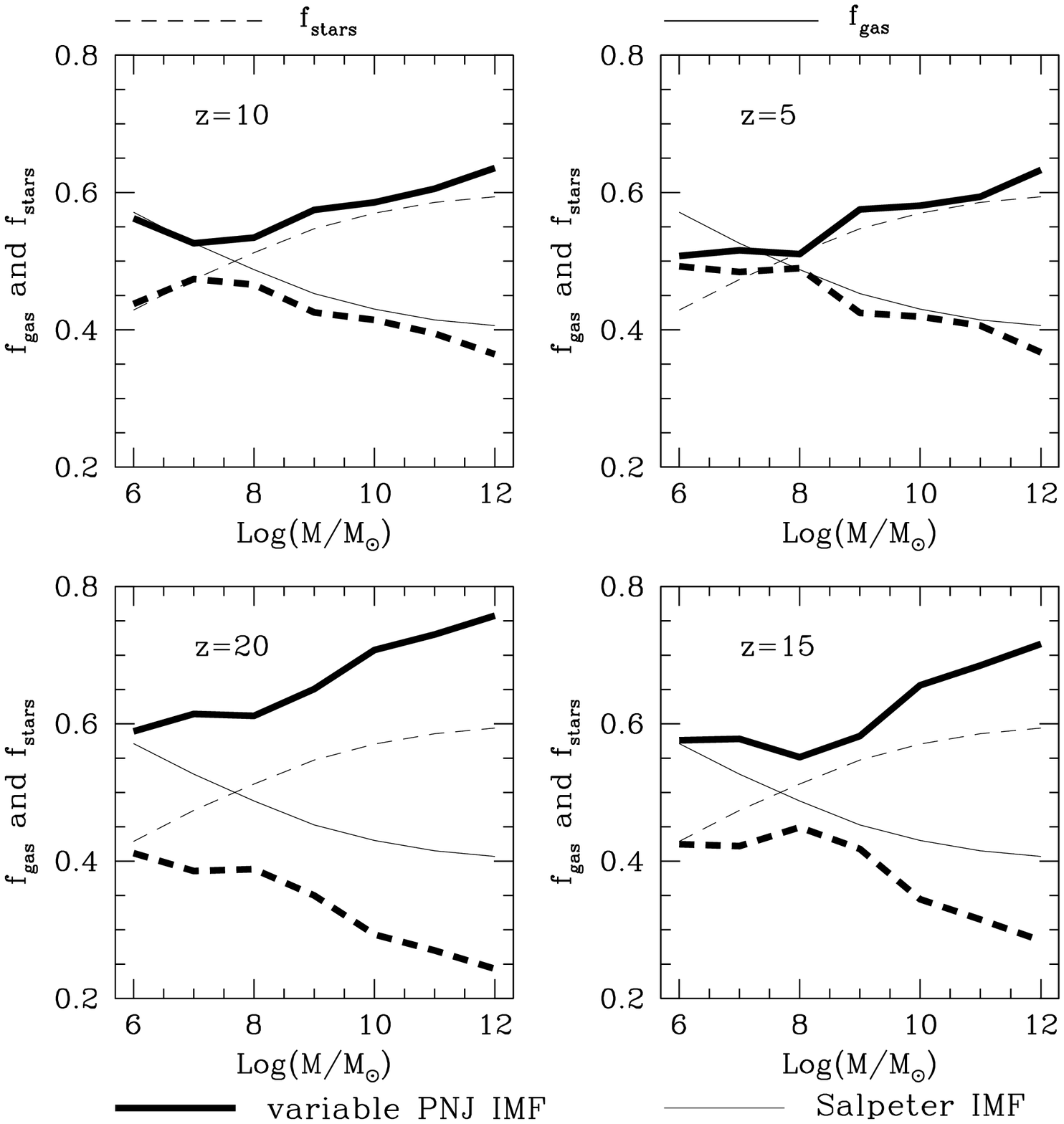,width=.56\textwidth} \hfil
\psfig{file=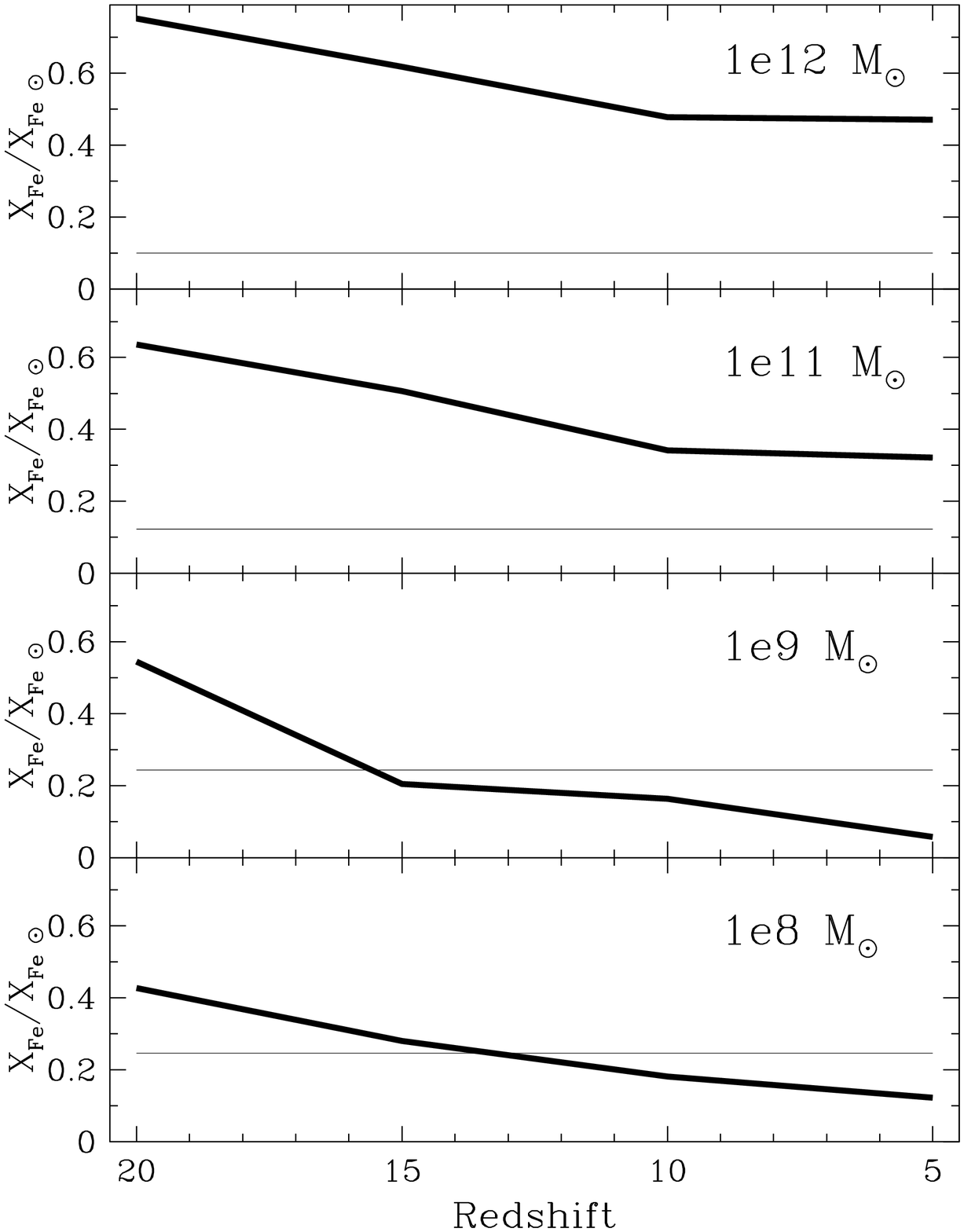,width=.42\textwidth}}
\caption{Comparison between galactic models with the variable PNJ IMF
({\it thick lines}) and models with the Salpeter IMF ({\it thin lines}).\\
{\it Left panels}: 
mass fraction of ejected gas ({\it solid lines}) and complementary
fraction locked into stars ({\it dashed lines}) as a function of the
initial  mass of the galaxy, for four different redshifts of formation 
as indicated. \\
{\it Rightmost panel}: metallicity of the gas ejected as GW
by galaxies of given initial baryonic mass, indicated in the 
plots, and as a function of their redshift of formation.}
\end{figure}

\noindent
In Fig.~1 (left panels) we compare the mass fraction of gas ejected in the GW,
and the complementary mass fraction locked into stars, for galactic models
with the variable PNJ IMF and for models with the Salpeter IMF
(thick and thin lines, respectively). Mass fractions refer to
the total initial baryonic mass of the galaxy. The amount
of ejected gas is larger in the case of the PNJ IMF, since in the early
galactic phases this IMF is skewed toward more massive stars and less
mass remains locked into long--lived, low-mass stars.
The difference with the Salpeter case gets sharper
for larger galactic masses, and for higher redshifts of formation.
Models with the Salpeter IMF evidently bear no dependence on the
redshift of formation.

The rightmost panel in Fig.~1 shows the iron abundances in the gas ejected 
as GW, again comparing the Salpeter IMF and the PNJ IMF case.
In most cases, the galactic ejecta in the PNJ models are more metal--rich 
than in the Salpeter case, up to a factor of 5 or more for the
more massive galaxies, and for high redshifts of formation. 
In the PNJ models, in fact, more gas in the galaxy gets recycled
through massive stars, effective metal contributors, while less gas
gets locked into low--mass stars.

From the trends described above, we expect galactic models 
with the PNJ IMF to predict, for the ICM, a more efficient metal pollution
and a higher fraction of the gas originating from GWs,
with respect to ``standard'' models. The first results in this respect
are discussed in Chiosi (2000).

\section{The chemical evolution of the ICM: a toy model}
Since the GW ejecta of ellipticals modelled with the PNJ IMF are sensitive
to the detailed redshift of formation of the individual galaxies,
to predict the chemical enrichment of the ICM we need to model 
the history of galaxy formation in the cluster.
To this aim, we developed a global, self-consistent
chemical model for the cluster, which can follow the
simultaneous evolution of all its components: the galaxies, the primordial
gas, and the gas processed and re-ejected via GWs (Moretti et~al.\ 2001).
Our chemical model for clusters is developed in analogy with the usual
chemical models for galaxies, as illustrated in the scheme below.

\begin{center}
\begin{large}
\begin{tabular}{c c c|c c c}
	    && ~~~~~\fbox{primordial gas}~~~~~	& 
~~~~~\fbox{primordial gas}~~~~~&& \\

	    && $\Downarrow$ 			& $\Downarrow$ && \\

& $\swarrow$ & {\normalsize SFR, IMF}		& {\normalsize GFR, GIMF} &
$\searrow$ & \\

	    && $\Downarrow$ 			& $\Downarrow$ && \\

ISM 	    && \fbox{stars}			& \fbox{galaxies} && ICM \\

	    && $\Downarrow$ 			& $\Downarrow$ && \\

& $\nwarrow$ & {\normalsize stellar yields} 	& {\normalsize GW yields} &
$\nearrow$ & \\

	    && $\Downarrow$ 			& $\Downarrow$ && \\

	    && \fbox{enriched gas}		& \fbox{enriched gas} && \\

\end{tabular}
\end{large}
\end{center}

\noindent
As the interstellar medium (ISM) is polluted by stars, the ICM
is polluted by galaxies.
The primordial gas in the ICM gets consumed in time by
some prescribed Galactic Formation Rate (GFR); at each time 
galaxies form distributed
in mass according to a Galactic Initial Mass Function (GIMF), derived from the 
Press-Schechter mass function suited to that redshift.
Through GWs, galaxies restitute chemically enriched gas,
which mixes with the overall ICM; the latter consists of the primordial 
gas not yet consumed by galaxy formation (if any) and of the gas re-ejected 
by galaxies up to the present age.

Model equation parallel those of galactic chemical models, with the 
substitutions {\mbox{SFR $\rightarrow$ GFR}}, 
{\mbox{IMF $\rightarrow$ Press-Schechter GIMF}}, 
{\mbox{stellar yields $\rightarrow$ GW yields}}. 
Model parameters are calibrated so that the resulting
galaxy formation history matches the observed present--day
luminosity function (LF) at the end of the simulation. 
For all details, see Moretti et~al.\ (2001).
\section{The ``best case'' models}
In Fig.~2 we show our case of ``best match'' with the observed LF 
in the B--band (Trentham 1998, top panels). The left panels refer 
to the case when galactic models with the PNJ IMF are adopted;
the right panels display results for the same cluster parameters
(i.e.\ same galaxy formation history), 
but adopting ellipticals with the Salpeter IMF.
The Salpeter case predicts somewhat more galaxies in the high--luminosity 
bins, due to the fact that for massive galaxies a larger mass fraction 
remains locked into stars in the Salpeter case than with the PNJ IMF 
(cf.\ Fig.~1). Anyways, the LF is still in agreement
with the observed one within errors.

\begin{figure}
{\centering 
\leavevmode
\psfig{file=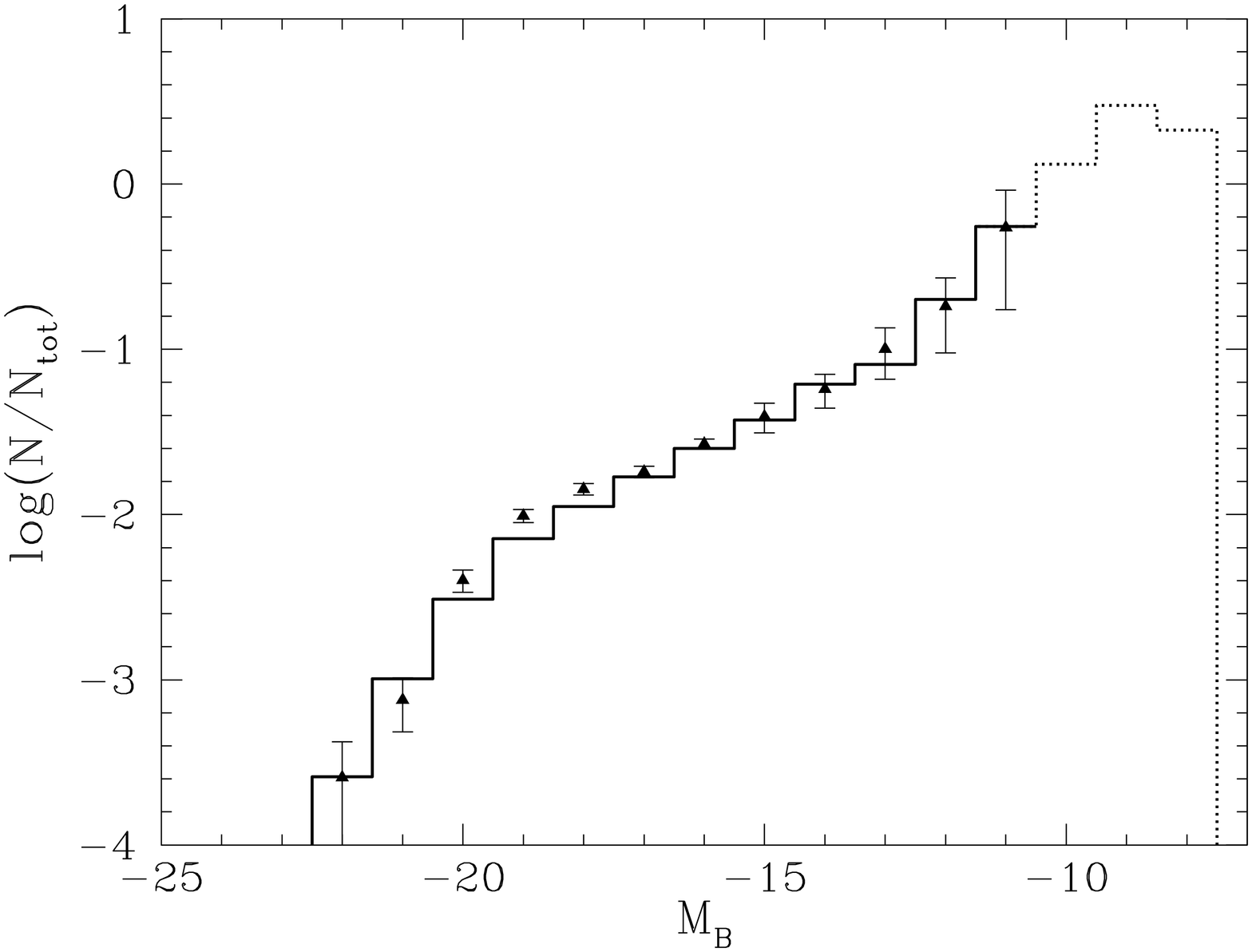,angle=-90,width=.49\textwidth} \hfil
\psfig{file=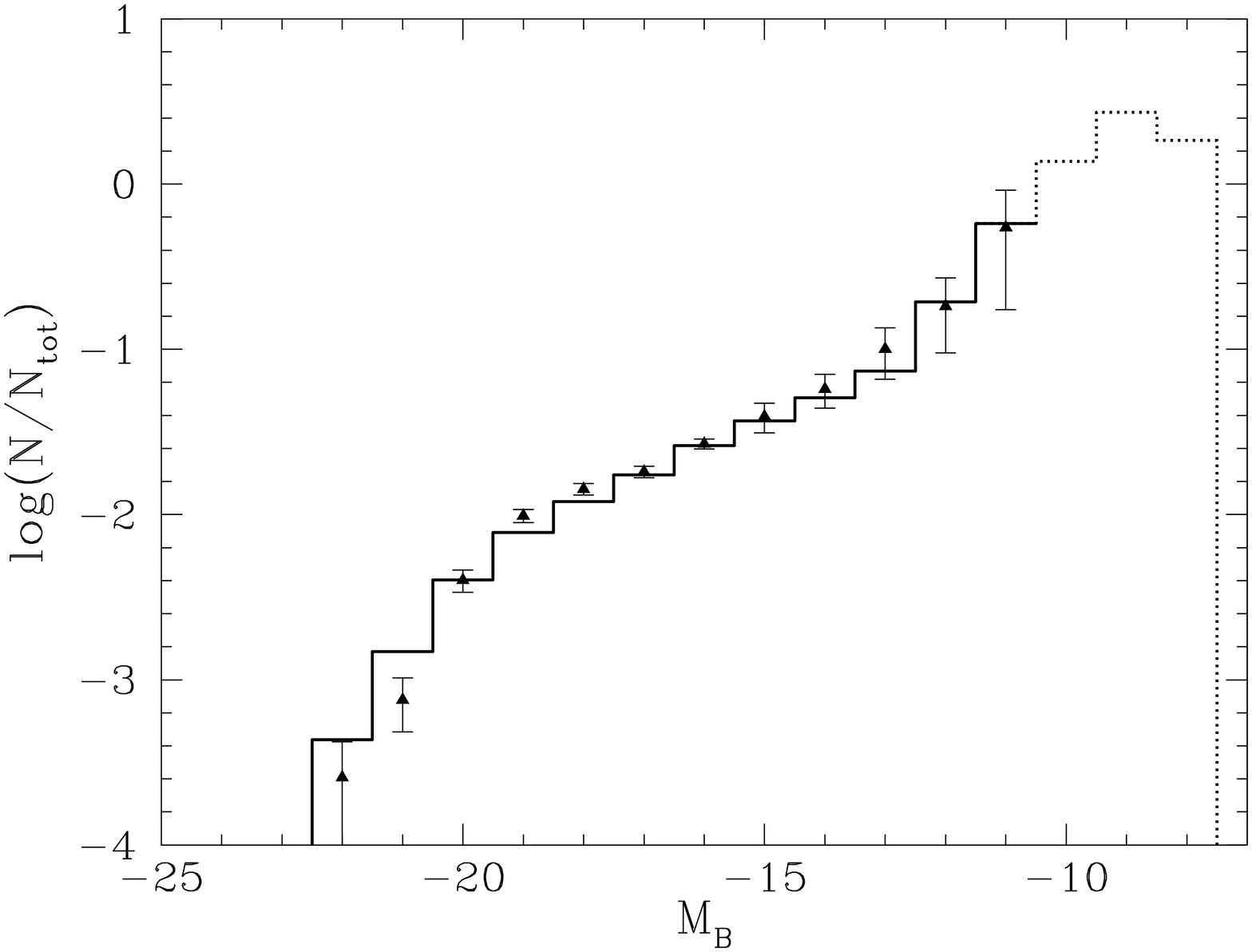,angle=-90,width=.49\textwidth}}
{\centering \leavevmode
\psfig{file=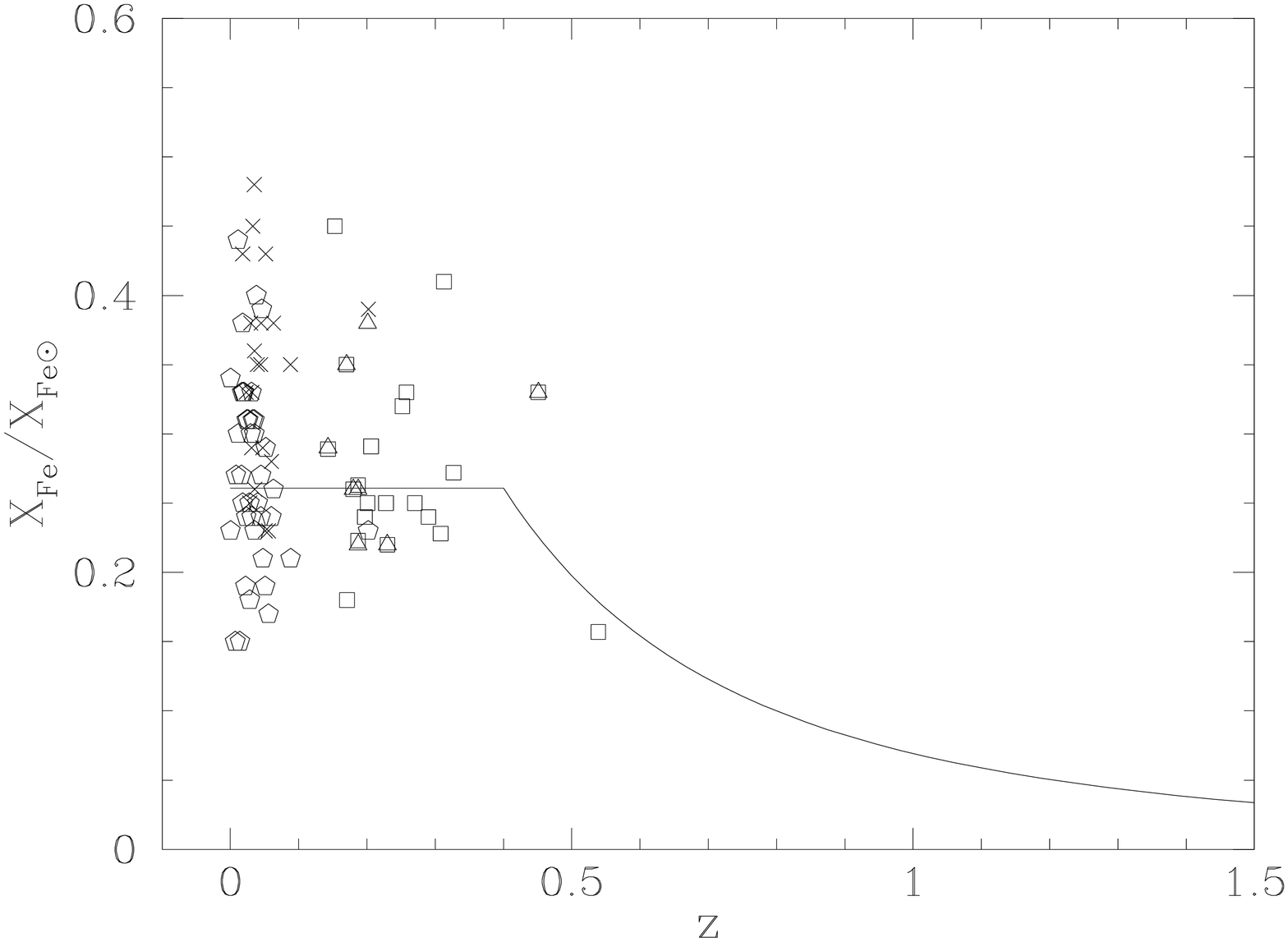,angle=-90,width=.49\textwidth} \hfil
\psfig{file=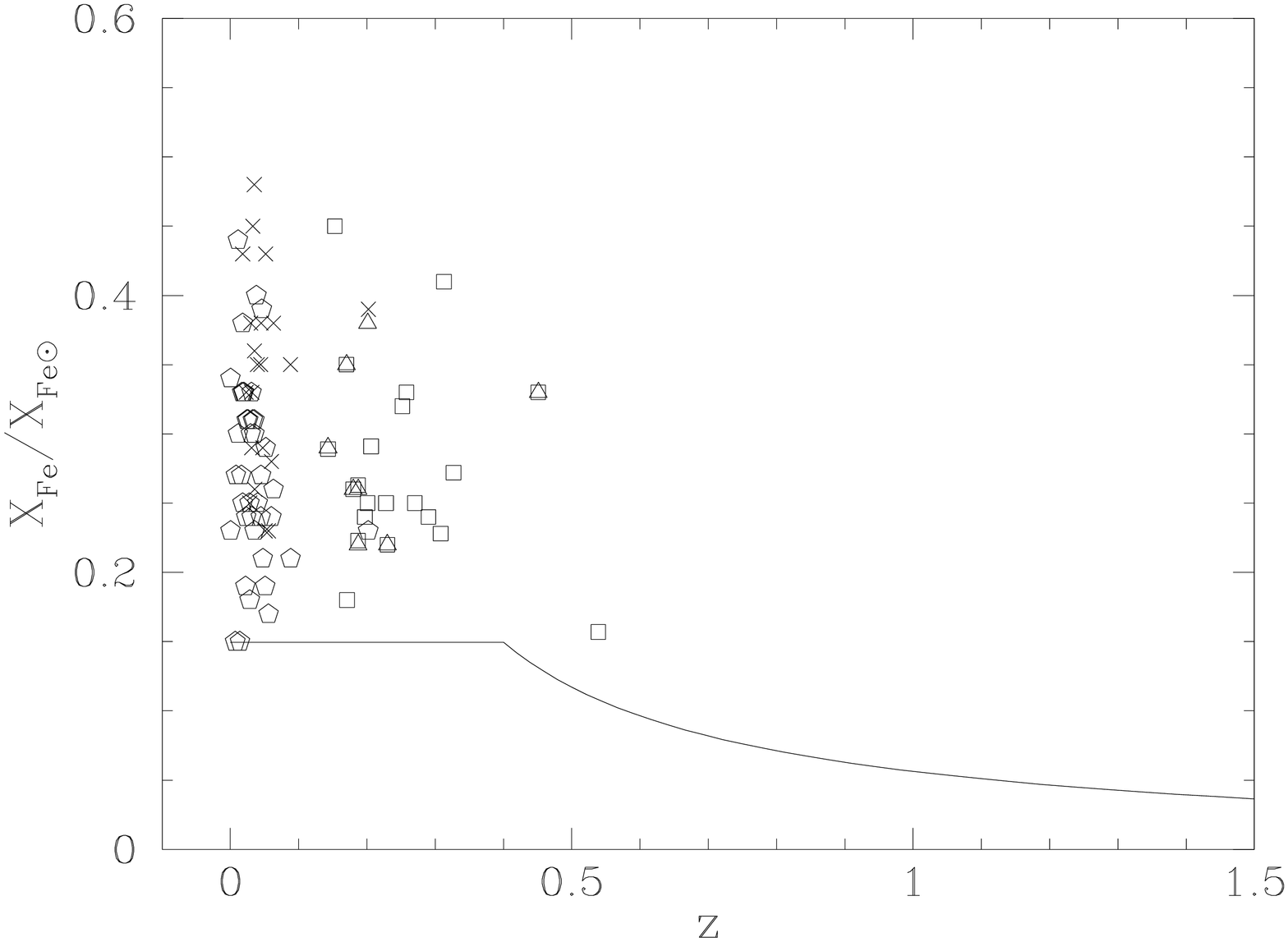,angle=-90,width=.49\textwidth}}
{\centering \leavevmode
\psfig{file=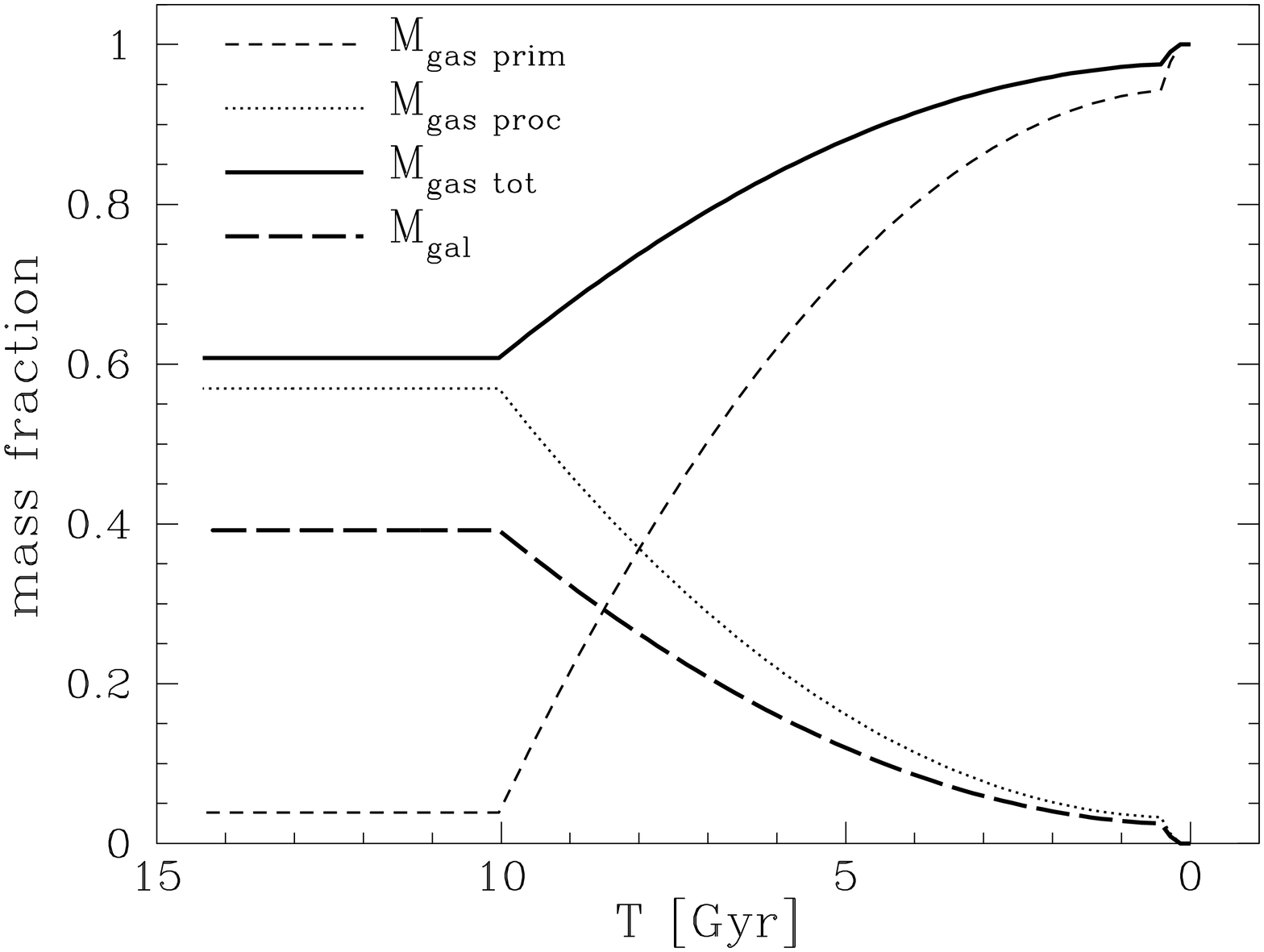,angle=-90,width=.49\textwidth} \hfil
\psfig{file=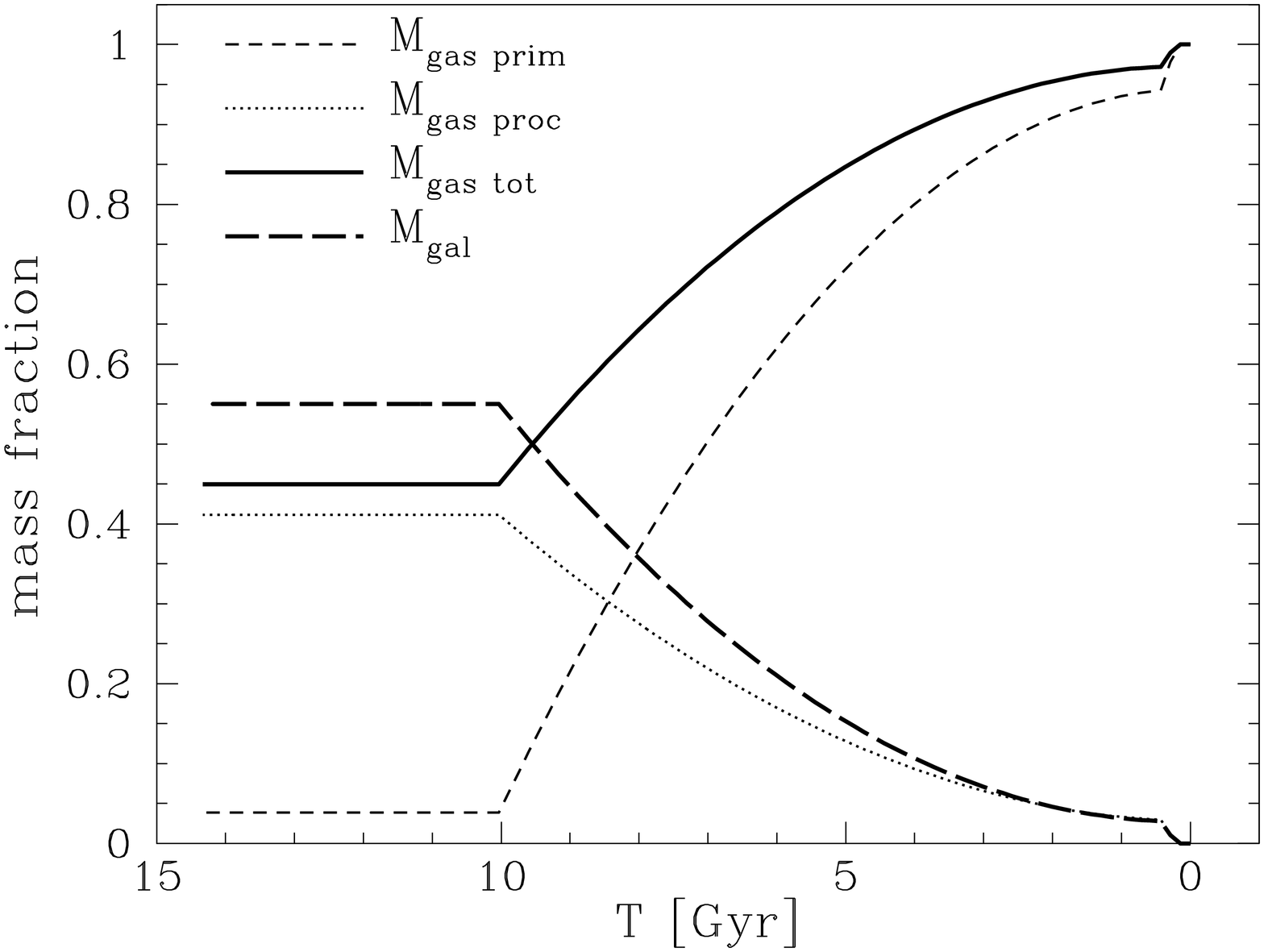,angle=-90,width=.49\textwidth}}
\caption{Results of our ``best case'' cluster model.
{\it Left panels}: results for
galactic models with the PNJ IMF; {\it right panels}: results for
galactic models with the Salpeter IMF.\\
{\it Top panels}:
B--band luminosity function of cluster galaxies versus the observational
data (by Trentham 1998). 
{\it Mid panels}:
predicted metallicity evolution of the ICM versus the observational
data (by Matsumoto et~al.\ 2000, Fukazawa et~al.\ 1998, 
Mushotzky \& Loewenstein 1997).
{\it Bottom panels}:
Time evolution of the cluster components: mass in primordial
and processed gas separately, total ICM gas mass, and mass in galaxies.
\label{fig:comparison}}
\end{figure}

Although the predicted LF is virtually the same in the two models,
strong differences are found in the predicted gas and metallicity content 
in the ICM. The mid panels in Fig.~2 show the predicted abundance
evolution in the ICM.
Adopting galactic models with the PNJ IMF clearly improves 
predictions about the metallicity of the ICM.

The bottom panels in Fig.~2 show the evolution of the mass fraction
of the various components of the cluster: the primordial gas, 
which gets consumed by galaxy formation; the processed gas, 
namely the gas that has been involved in galaxy formation
and then re-ejected as GW; the total gas, sum of the primordial 
and of the processed gas; the mass in galaxies, that is in the stellar
component we see today, ``left over'' after the GW. While in the Salpeter
case (right panel) the overall mass that remains locked into galaxies 
(long--dashed line) is larger that the mass ejected in the GWs 
(dotted line), the opposite is true for the cluster model 
with the PNJ galaxies (left panel), as qualitatively expected from \S~2.
In the latter case, the mass of the re-ejected
gas is $\sim 1.5$ times larger than that locked into galaxies. Although
this is not enough to account for the whole of the observed intra-cluster gas 
(whith a mass 2--5 times larger than that in galaxies, Arnaud et~al.\ 1992), 
the amount of gas re-ejected by galaxies is expected
to make up for a remarkable fraction of the overall ICM.

\section{Open issues and future perspectives}
Once the model is calibrated to reproduce the observed LF in the B--band
(Fig.~2, top panels), it turns out that the match with LFs in redder bands 
is not as good. Fig.~3 (left panel) shows the comparison to the observed LF
in the R--band: the ``best--case'' model calibrated on the B--band seems to
underestimate the number of luminous galaxies with a red stellar population.
A similar effect is seen for the LF in the K--band. We used
the B--band LF for the calibration, as it offered the deepest and most 
extensive dataset, but the LF in the red bands is probably a better track
of the old stellar population responsible for the bulk of the metal enrichment,
while the B--band might be sensitive to recent minor bursts of star formation.
Hence, calibrating the model over the red stellar population should provide
a better estimate of the galaxy formation history and of the consequent
chemical enrichment of the ICM.

In particular, a larger number of old giant galaxies will help
to obtain higher values of the Iron Mass to Light Ratio
(IMLR, for a definition see Renzini 1997 and references therein),
closer to the very high IMLRs measured in real clusters ($\geq 0.01$, 
Finoguenov, David \& Ponman 2000).
To illustrate this point, in the right panel of Fig.~3 we plot 
the present--day IMLR of individual ellipticals modelled with the PNJ IMF, 
as a function of the initial
mass of the galaxy and for different redshifts of formation. 
Higher values of the IMLR pertain to more massive and older galaxies.
With the PNJ IMF in fact, galaxies which are more massive and/or formed 
at higher redshifts, store less mass in the stellar component, while ejecting 
more, and more metal--rich, gas in the GW (Fig.~1 and \S~2); 
both effects tend to enhance the corresponding IMLR.

Hence, with a galaxy formation history producing more red giant galaxies
in the ``cluster mixture'', as suggested by the LFs in the red bands,
we expect to predict high values for the overall IMLR in the cluster.

\begin{figure}[t]
{\centering 
\leavevmode
\psfig{file=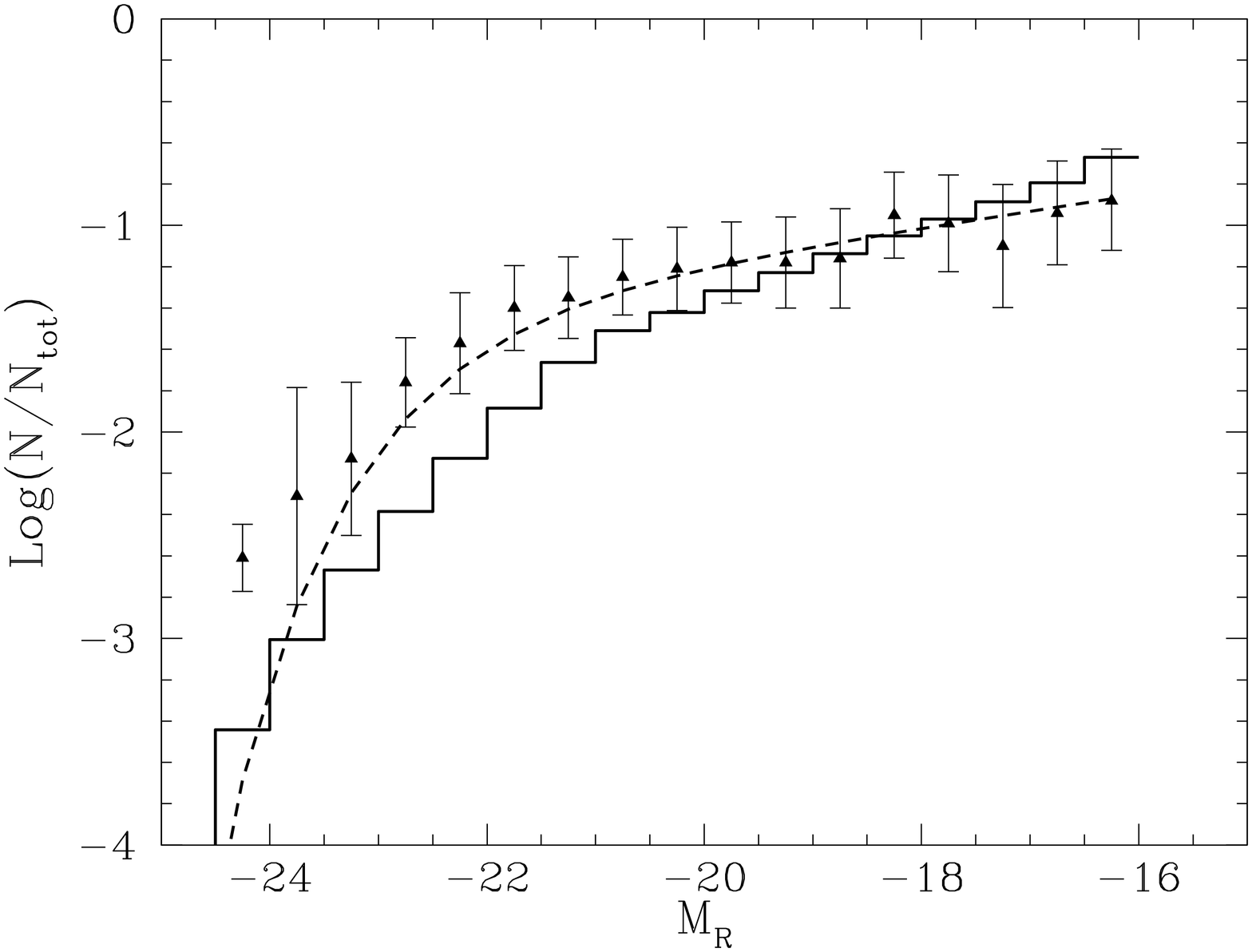,angle=-90,width=.49\textwidth} \hfil
\psfig{file=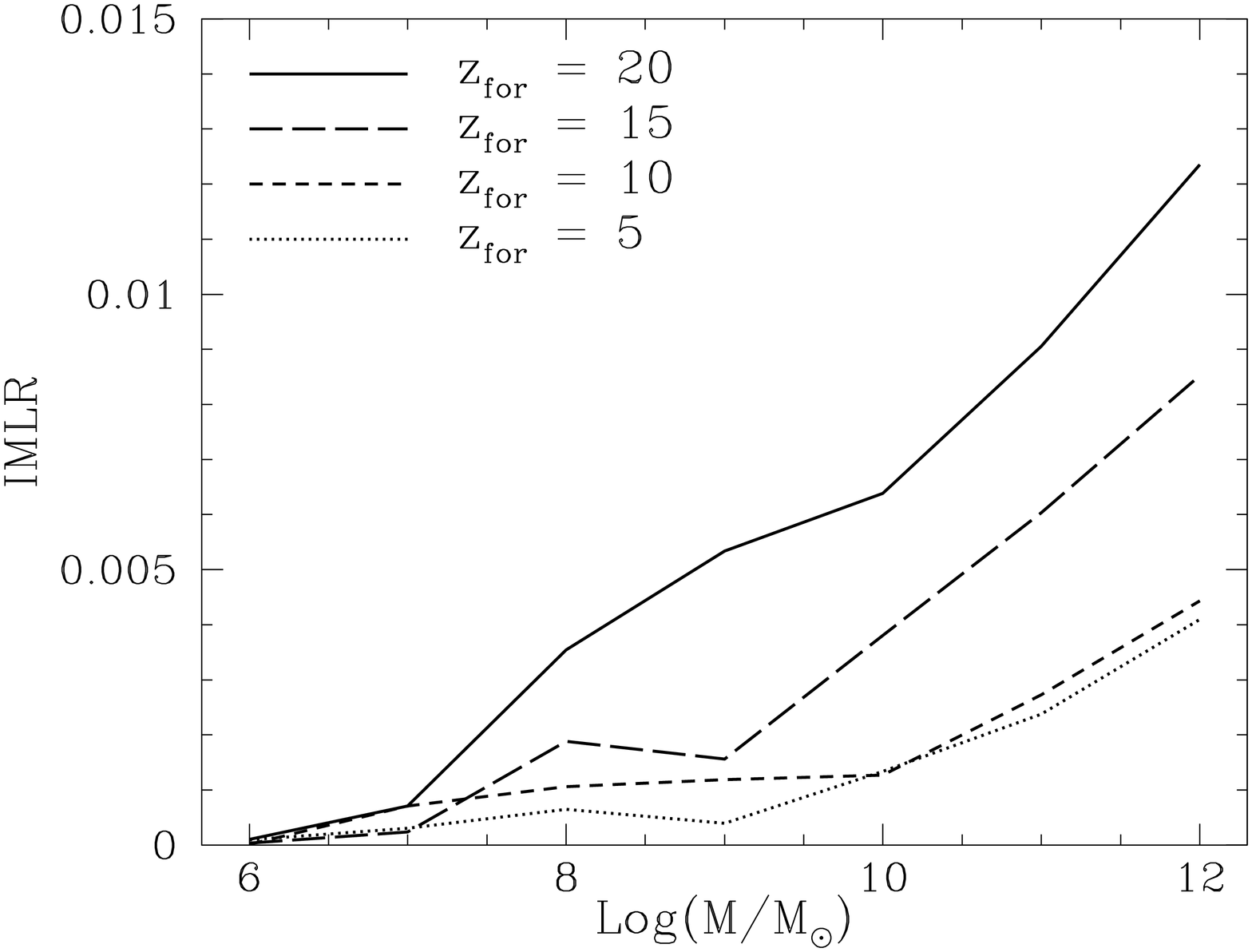,angle=-90,width=.49\textwidth}}
\caption{ {\it Left panel}: 
Predicted R--band luminosity function of cluster galaxies
versus the observational data (by Driver et~al.\ 1998; the dashed line is the 
Schechter LF they derive from their data).\\
{\it Right panel}: Iron Mass to Light Ratio for individual ellipticals
as a function of initial mass and for different redshifts of formation.}
\end{figure}

\section*{Acknowledgments}
\vspace{-3mm}
LP and AM are grateful to F.~Matteucci and to the organizers of this conference
for giving them the opportunity to participate and contribute.


\end{document}